\documentclass[journal,twocolumn,10pt]{IEEEtran}
\usepackage{amssymb}

\usepackage{amsfonts, amsmath, amssymb}
\usepackage{commands11}
\usepackage{cite}
\usepackage{color}
\usepackage{subfigure}
\usepackage{graphicx}
\usepackage{algorithm}
\usepackage{algorithmic}
\usepackage{verbatim}

\renewcommand{\B}[1]{\boldsymbol{#1}}
\newcommand{\ind}[1]{1\hspace{-2mm}{1}\left(#1\right)}
\newcommand{\Probt}[2]{\Prob_{\text{#1}}^{\text{#2}}}

\newtheorem{proposition}{Proposition}

\allowdisplaybreaks
\IEEEoverridecommandlockouts

\newcommand{\LZ}[1]{{\color{black} #1}}
\newcommand{\DG}[1]{{\color{black} #1}}
\newcommand{\DGnew}[1]{{\color{black} #1}}
\newcommand{\LZnew}[1]{{\color{black} #1}}
\newcommand{\LZH}[1]{{\color{black} #1}}

\begin{document}
\title{Virtual Full Duplex Wireless Broadcast\DGnew{ing} via \DGnew{Compressed Sensing}}

\author{
Lei Zhang and Dongning Guo\\
Department of Electrical Engineering \& Computer Science \\
Northwestern University,
Evanston, IL 60208, USA
\thanks{ This work has been presented in part at the $2011$ IEEE International Symposium on Information Theory~\cite{ZhaGuo11ISIT}.}
}

\maketitle

\begin{abstract}
A novel solution is proposed to undertake a frequent task in wireless networks,
which is to let all nodes broadcast information to and receive information from their respective one-hop neighboring nodes.
\DGnew{The contribution is two-fold.  First, as each neighbor selects one message-bearing codeword from its unique codebook for transmission, it is shown that decoding their messages based on a superposition of those codewords through the multiaccess channel is fundamentally a problem of compressed sensing.}
In the case where each message consists of a small number of bits, an iterative 
algorithm based on belief propagation is developed for \DGnew{efficient decoding}.
\DGnew{Second, to satisfy the half-duplex constraint,
each
codeword
consists of randomly distributed} on-slots and off-slots.
A node transmits during its on-slots, and listens to its neighbors
only through its own off-slots.
\DGnew{Over one frame interval, each node broadcasts a message to neighbors and simultaneously}
decodes neighbors' messages 
based on the superposed signals received through its own off-slots.
Thus the solution fully exploits the multiaccess nature of the wireless medium and addresses the half-duplex constraint at the fundamental level.
In a network consisting of Poisson distributed nodes, numerical results demonstrate that the proposed scheme often achieves several times the rate of slotted ALOHA and CSMA with the same packet error rate.
\end{abstract}

\section{Introduction} \label{sec:intro}
Consider a frequent situation in wireless peer-to-peer networks, where every node wishes to broadcast messages to all nodes within its one-hop neighborhood, called its {\em neighbors}, and also wishes to receive messages from its neighbors.  We refer to this problem as {\em mutual broadcasting}.  Such traffic can be dominant in many applications, such as messaging or video conferencing of multiple parties in a spontaneous social network or at an incident scene. Wireless mutual broadcasting is also critical to efficient network resource allocation, where messages are exchanged between nodes about their local states, such as queue length, channel quality, code and modulation format, and request for certain resources and services.

A major challenge in wireless networks is the half-duplex constraint,
namely, currently affordable radio cannot receive useful signals at the same time over the same frequency band over which it is transmitting. This is largely due to the limited dynamic range and noise of the radio frequency circuits, which are likely to remain a physical restriction in the near future. An important consequence of the half-duplex constraint is that, if two neighbors transmit their packets \LZ{(or frames which are used interchangeably hereafter)} at the same time, they do not hear each other. To achieve reliable mutual broadcasting using a usual packet-based scheme, nodes have to repeat their packets many times interleaved with random delays, so that all neighbors can hear each other after enough retransmissions. This is basically the ubiquitous random channel access solution.

A closer examination of the half-duplex constraint, however, reveals
that a node does not need to transmit an entire packet before
listening to the channel. An alternative solution is conceivable: Let
a frame (typically of a few thousand symbols) be divided into some
number of slots, where a node transmits over a subset of the slots
and assumes silence over the remaining slots, then the node can receive useful signals over those nontransmission slots. If different nodes activate different sets of on-slots, then \LZ{they} can all transmit information during a frame and receive useful signals within the same frame, and decode messages from neighbors as long as sufficiently strong error-control codes are applied.

This on-off signaling, called {\em rapid on-off-division duplex (RODD)}, was originally proposed in~\cite{GuoZha10Allerton}. Using RODD, reliable mutual broadcasting can be achieved using a single frame interval. Thus RODD enables half-duplex radios to achieve virtual full-duplex communication. Despite the half-duplex physical layer, the radio appears to be full-duplex in higher layers. Not only is RODD signaling applicable to the mutual broadcasting problem, it can also be the basis of a clean-slate design of the physical and medium access control (MAC) layers of wireless peer-to-peer networks.

In this paper, we focus on a special use of RODD signaling and a
special case of mutual broadcasting, where every node has a small number
of bits to send to its neighbors. It is assumed that node transmissions are perfectly synchronized. The goal here is to provide a practical
algorithm for encoding and decoding the short messages to achieve
reliable and efficient mutual broadcasting. Decoding is
fundamentally 
a problem
of {\em sparse recovery} \DGnew{(or {\em compressed sensing})}
based on linear measurements, since the received
signal is basically a noisy superposition of neighbors' codewords selected
from their respective codebooks. There are many algorithms developed
in the compressed sensing
literature to
solve the problem, the complexity of which is often polynomial in the
size of the codebook~(see, e.g.,
\cite{NeeTro09ACHA,DonMal10ITW,Maleki11Thesis,DaiMil09IT, BarSar10TSP}). In this
paper, an iterative message-passing algorithm based on belief
propagation (BP) with linear complexity is developed.
Numerical results show that the proposed RODD scheme significantly outperforms slotted-ALOHA with multi-packet reception capability and CSMA in terms of data rate.

The excellent performance of the proposed scheme is because it departs from the usual solution where a highly reliable, highly redundant, capacity-achieving, point-to-point physical-layer code is paired with a rather unreliable MAC layer.  By treating the physical and MAC layers as a whole, the proposed scheme achieves better overall reliability at much higher efficiency.

It is easier to implement RODD in a synchronized network.
Regardless of whether RODD or any other physical- and MAC-layer technology is used,
it is necessary to acquire their timing (or relative delay) in order to decode messages from neighbors.
 Timing acquisition and decoding are generally easier if the frames arriving at a receiver are synchronous locally within each neighborhood, although synchronization is not a necessity. In a wireless network, synchronization can be achieved using various distributed algorithms for reaching consensus~\cite{SchRib08TSP, SchGia08TSP, SimSpa08SPM} or using a common source of timing, such as the Global Positioning System (GPS). Whether synchronizing the nodes is worthwhile is a challenging question, which is not discussed further in this paper.

\DGnew{In contrast to virtual full duplex using RODD signaling, real \LZH{full duplex} 
becomes feasible if the interference a node's transmit chain causes its own receive chain can be suppressed to weaker level than the desired received signal.
Two self-interference cancellation techniques have recently received much attention:
One employs a balanced/unbalanced transformer to negate the transmitted signal for analog
cancellation~\cite{JaiCho11Mobicom};
the other 
separates the transmit and receive antennas and uses analog and optional digital cancellation~\cite{SahPat11X}.
However, those techniques do not always apply.
For instance, self-interference cancellation may be insufficient if the signals have large dynamic range;
space limitations may not allow for adequate antenna separation; and,
with multiple transmit antennas, canceling self interference in multiple chains may be hard. In those cases, RODD is a more viable solution.
Whether full duplex is achieved using RODD or an alternative means, the proposed compressed sensing framework and technique provide a competitive solution to mutual broadcasting.
}

 The remainder of the paper is organized as follows. \LZ{The system model is presented in Section~\ref{sec:Model}.  The proposed coding scheme for mutual broadcasting is described in Section~\ref{sec:EncMB}. The message-passing decoding algorithm is developed in Section~\ref{sec:DecMB}. Section~\ref{sec:Compare} studies the conventional random access schemes, namely slotted-ALOHA with multi-packet reception capability and CSMA.} Numerical comparisons are presented in Section~\ref{sec:nr}. Section~\ref{sec:Conclude} concludes the paper.

\section{The System Model} \label{sec:Model}

\subsection{Linear Channel Model}
Let $\Phi=\{Z_i\}_i$ denote the set of nodes on the plane. We refer to a node by its location $Z_i$. Suppose all transmissions use the same single carrier frequency.
\footnote{\DGnew{The frequency offset between different transmitters is assumed to be small, so as not to cause phase rotation over one frame.}}  Let time be slotted and all nodes be perfectly synchronized.\footnote{A discussion of
synchronization issues \DGnew{is found in~\cite{GuoZha10Allerton}}. In~\cite{AppBaj12PhyComm}, cyclic codes are proposed to \DGnew{accommodate different}
user delays \DGnew{under the same compressed sensing framework}.
} \LZnew{Suppose each node has $l$ bits to broadcast to its neighbors.} Let $d_i\in\{1,\dots,2^l\}$ denote the data or message node $Z_i$ wishes to broadcast. In discrete-time baseband, let $\B{S}_i(d_i)$ denote the \LZ{on-off} signature (codeword) of length $M_s$ transmitted by node $Z_i$, whose entries take values in $\{-1,0,+1\}$. \LZ{Here, the zero entries of $\B{S}_i(d_i)$ correspond to the off-slots in which $Z_i$ listens to the channel. The design of the on-off signatures will be discussed in Section~\ref{sec:EncMB}.} Let $U_{0i}$ denote the complex-valued coefficient of the wireless link from $Z_i$ to $Z_0$. The signal received by node $Z_0$, if it could listen over the entire frame, is described by
\begin{equation} \label{eq:PhyModel}
\widetilde{\B{Y}} = \sqrt{\gamma} \sum_{Z_i \in \Phi\backslash\{Z_0\}} U_{0i}\B{S}_i(d_i)+ \widetilde{\B{W}}
\end{equation}
where \DGnew{the noise} $\widetilde{\B{W}}$ 
\DGnew{consists} of independent identically distributed (i.i.d.) circularly symmetric complex Gaussian entries with zero mean and unit variance, and $\gamma$ denotes the nominal signal-to-noise ratio (SNR). Denote the set of neighbors of $Z_0$ by $\mathcal{N}(Z_0)$.
If we further assume that transmissions from non-neighbors, if any, are accounted for as part of the additive Gaussian noise, \eqref{eq:PhyModel} can be rewritten as
\begin{equation} \label{eq:PhyModel2}
\widetilde{\B{Y}} = \sqrt{\gamma} \sum_{Z_i \in \mathcal{N}(Z_0)} U_{0i}\B{S}_i(d_i)+\overline{\B{W}}
\end{equation}
where each element in $\overline{\B{W}}$ is assumed to be circularly symmetric complex Gaussian with variance $\sigma^2$.
The variance, to be derived in Section~\ref{sec:EncMB}, accounts for interference from non-neighbors and depends on the network topology.

\subsection{Network Model} \label{sec:network}
Consider a network with nodes distributed across the plane according to a homogeneous Poisson point process (p.p.p.) with intensity $\lambda$. The number of nodes in any region of area $A$ is a Poisson random variable with mean $\lambda A$. Without loss of generality, we assume node $Z_0$ is located at the origin and focus on its performance, which should be representative of any node in the network.

Poisson point process is the most frequently used model to study wireless networks (see~\cite{BacBla09FT} and references therein).
The homogeneous p.p.p.\ model is assumed here to facilitate analysis and comparison of competing technologies.
The proposed RODD signaling and the mutual broadcasting scheme are not limited to Poisson distributed networks.

\subsection{Propagation Model and Neighborhood} \label{sec:neighbor}

The large-scale signal attenuation over distance 
is assumed to follow the power law with some path-loss exponent $\alpha>2$.  The small-scale fading of a link is modeled by an \DG{independent} Rayleigh random variable with mean equal to $1$.
The neighborhood of a node can be defined in many different ways.
For concreteness, we say that nodes $Z_i$ and $Z_j$ are neighbors of each other if the channel gain between them 
exceeds a certain threshold, 
$\theta$. Link reciprocity is regarded as given.

For any pair of nodes $Z_i,Z_j\in\Phi$, let $R_{ij}=|Z_i-Z_j|$
denote the distance and
and $G_{ij}$ the small-scale fading gain between them in a given frame, respectively. Then the channel gain between $Z_j$ and $Z_i$ is $G_{ij}R_{ij}^{-\alpha}$. The neighborhood of a node depends on the instantaneous fading gains. Specifically, we denote the set of neighbors of node $Z_i$ as
\begin{equation}
\mathcal{N}(Z_i) = \left\{Z_j\in\Phi: G_{ij}R_{ij}^{-\alpha}\geq \theta,j\neq i\right\}.
\end{equation}
The channel coefficient $U_{ij}$ should satisfy $|U_{ij}|^2=G_{ij}R^{-\alpha}_{ij}$, where its phase is assumed to be uniformly distributed on $[0,2\pi)$ independent of everything else.  Assuming the Poisson point process network model introduced in Section~\ref{sec:network}, the distribution of the amplitude of coefficient $U_{0i}$ in~\eqref{eq:PhyModel} for an arbitrary neighbor $Z_i\in \mathcal{N}(Z_0)$ is derived in the following.

Without loss of generality, we drop the indices $0$ and $i$, and use $R$ and $G$ to denote the distance and the fading gain, respectively. Since the two nodes are assumed to be neighbors, $G$ and $R$ satisfy $GR^{-\alpha}\geq\theta$, i.e., $R\leq\left(G/\theta\right)^{1/\alpha}$. Under the assumption that all nodes form a p.p.p., for given $G$, this arbitrary neighbor $Z_i$ is uniformly distributed in a disc centered at node $Z_0$ with radius $\left(G/\theta\right)^{1/\alpha}$. Therefore, the conditional distribution of $R$ given $G$ can be expressed as
\begin{align} \label{eq:condProb}
\Prob(R\leq r\big|G) = \min\left\{1,\left(\frac{\theta}{G}\right)^\frac{2}{\alpha} r^2\right\}.
\end{align}
Hence, for every $u \geq \sqrt{\theta}$, 
\begin{align}
\Prob(GR^{-\alpha}\geq u^2) &= \expsub{G}{\Prob\left(R \leq \left(\frac{G}{u^2}\right)^{\frac{1}{\alpha}} \bigg|G \right)} \nonumber \\
&= \expsub{G}{\left(\frac{G}{u^2}\right)^\frac{2}{\alpha} \left(\frac{\theta}{G}\right)^\frac{2}{\alpha}} \nonumber \\
&= \frac{\theta^{\frac2\alpha}}{u^{\frac4\alpha}}\,. \label{eq:ucdf}
\end{align}
Therefore, the probability density function (pdf) of $|U_{0i}|$ 
is
\begin{align} \label{eq:updf}
  p(u) = \left\{
    \begin{array}{ll}
      \frac{4}{\alpha} \frac{\theta^{2/\alpha}}{u^{4/\alpha+1}}, & u \geq \sqrt{\theta}; \\
      0, & \text{otherwise}.
    \end{array}
    \right.
\end{align}

In fact, the coefficient vector $\mathcal{G}_i=(G_{ji})_j$ for all $j\ne i$ can be regarded as a mark of node $Z_i$, so that $\tilde{\Phi}=\{(Z_i,\mathcal{G}_i)\}_i$ is a marked p.p.p.~\cite{Kingma93}.  Denote
\begin{align} \label{eq:Phihat}
\hat{\Phi}=\tilde{\Phi}\backslash (Z_0,\mathcal{G}_0)
\end{align}
given that $(Z_0,\mathcal{G}_0)$ is at the origin. By the Slivnyak-Meche theorem~\cite{BacBla09FT}, $\hat{\Phi}$ is also a marked p.p.p. with intensity $\lambda$. By the Campbell's theorem~\cite{BacBla09FT}, the average number of neighbors of $Z_0$ can be obtained as:
\begin{align}
c &= \expsub{\hat{\Phi}}{\sum_{(Z_i,\mathcal{G}_i)\in\hat{\Phi}} \ind{G_{0i}R_{0i}^{-\alpha}\geq \theta}} \nonumber \\
&= 2\pi\lambda\int_0^\infty\int_0^\infty \ind{gr^{-\alpha}\geq \theta}re^{-g}\diff r \diff g \nonumber \\
&= \frac{2}{\alpha}\pi\lambda\theta^{-2/\alpha}\Gamma\left(\frac{2}{\alpha}\right) \label{eq:c}
\end{align}
where $\ind{\cdot}$ is the indicator function and $\Gamma(\cdot)$ is the Gamma function.

\section{Encoding for Mutual Broadcasting}\label{sec:EncMB}

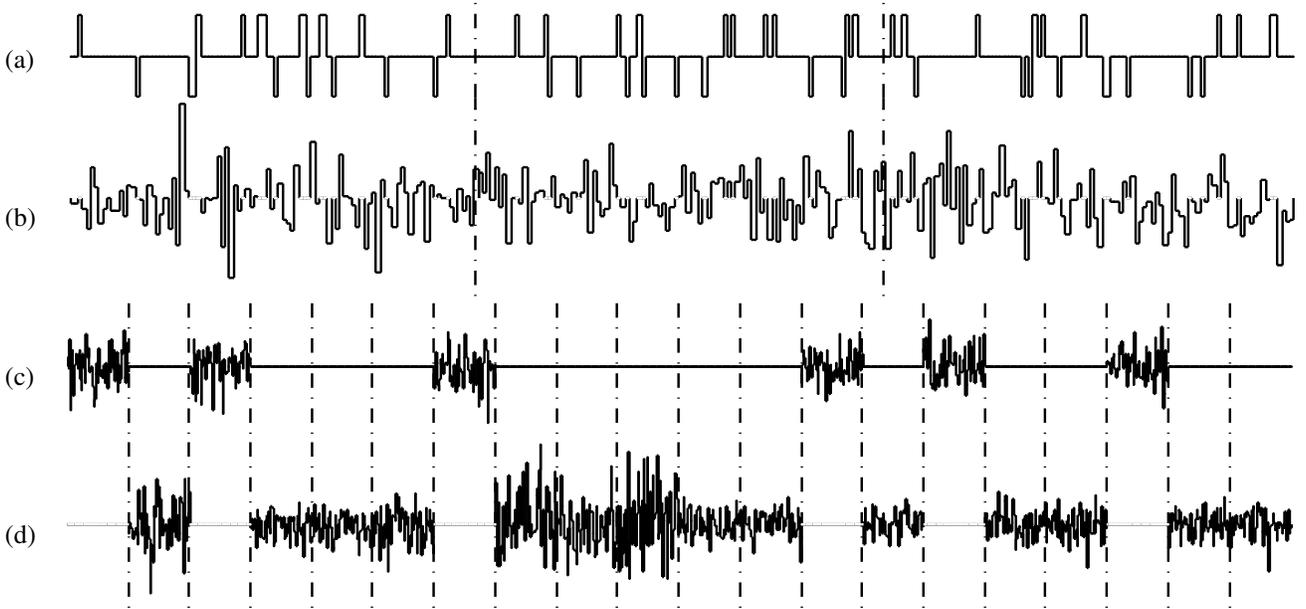
\begin{figure*}
\centering
\begin{picture}(20,225)
  \put(0,205){(a)}
  \put(0,145){(b)}
  \put(0,85){(c)}
  \put(0,25){(d)}
\end{picture}
\includegraphics[width=.9\textwidth]{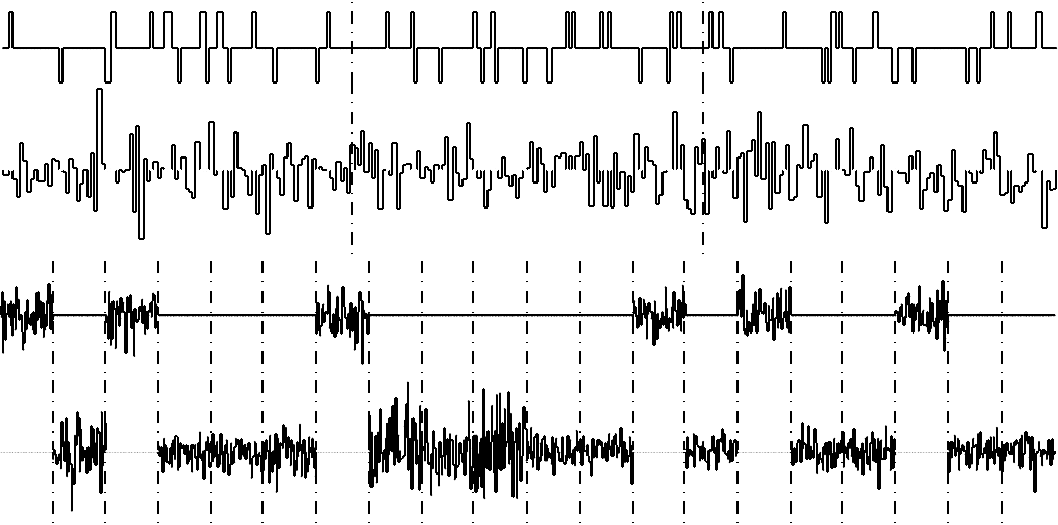}
\caption{Signals a node transmits and receives using RODD and slotted-ALOHA, respectively:
(a) RODD signaling, three consecutive frame transmissions by a node;
(b) RODD signaling, three frames received by the same node (blanks represent erasures);
(c) slotted-ALOHA, transmissions of a node over 20 frame intervals;
and
(d) slotted-ALOHA, signals received by the same node over 20 frame intervals.  All signals are of the same symbol rate and bandwidth.
The received samples are complex-valued due to the complex-valued channel gain.  We only plot the in-phase component of the received signals.
}
\label{fig:signals}
\end{figure*}

Recall that each node $Z_i$ is assigned a codebook of $2^l$ on-off signatures (codewords) of length $M_s$, denoted by $\{\B{S}_i(1), \dots, \B{S}_i(2^l)\}$.
 For simplicity, let each element of each signature be generated randomly and independently, which is $0$ with probability $1-q$, and $1$ and $-1$ with probability $q/2$ each.\footnote{The optimal design of the signatures is out of the scope of this paper.} Node $Z_i$ broadcasts its $l$-bit message (or information index) $d_i \in \{1,\dots,2^l\}$ by transmitting the codeword $\B{S}_i(d_i)$.
Transmissions of all nodes are synchronized.
 \LZ{All nodes finish one message exchange after each frame of transmission.}

In each symbol slot, those transmitting nodes in $\hat{\Phi}$ \LZ{defined in~\eqref{eq:Phihat}} form an independent thinning of $\hat{\Phi}$ with retention probability $q$, denoted by $\hat{\Phi}_q$. $\hat{\Phi}_q$ is still an independent marked p.p.p., but with intensity $\lambda q$. Thus, the sum power from all transmitting non-neighbors of node $Z_0$ in each time slot is derived as
\begin{align}
&\expsub{\hat{\Phi}_q}{\sum_{(Z_i,\mathcal{G}_i)\in\hat{\Phi}_q} \gamma G_{0i}R_{0i}^{-\alpha}\ind{G_{0i}R_{0i}^{-\alpha}<\theta}} \nonumber \\
&\quad= 2\pi\lambda q\gamma \int_0^\infty\int_0^\infty gr^{-\alpha} \ind{gr^{-\alpha}<\theta}re^{-g}\diff r \diff g \nonumber \\
&\quad= 2\pi\lambda q\gamma \int_0^\infty r^{-\alpha+1}\left[1-(\theta r^\alpha+1) e^{-\theta r^\alpha}\right] \diff r \nonumber \\
&\quad= \frac{4\pi\lambda q\gamma\theta}{\alpha-2}\int_0^\infty re^{-\theta r^\alpha} \diff r \nonumber \\
&\quad= \frac{4}{\alpha(\alpha-2)}\pi\lambda q\gamma\theta^{1-2/\alpha} \Gamma\left(\frac{2}{\alpha}\right)\,.
\end{align}
Therefore, the variance of each element of $\overline{\B{W}}$ in~\eqref{eq:PhyModel2} is
\begin{equation}
\sigma^2=\frac{4}{\alpha(\alpha-2)}\pi\lambda q\gamma\theta^{1-2/\alpha} \Gamma\left(\frac{2}{\alpha}\right)+1.
\end{equation}

The signal received by the typical node $Z_0$, if it could listen over the entire frame, is described by~\eqref{eq:PhyModel2}. Suppose $|\mathcal{N}(Z_0)|=K$ and the neighbors of $Z_0$ are indexed by $1,2,\dots,K$. The total number of signatures of all neighbors is $N=2^lK$.  Due to the half-duplex constraint, however, node $Z_0$ can only listen during its off-slots, the number of which has binomial distribution, denoted by $M\sim \mathcal{B}(M_s,1-q)$, whose expected value is $\expect{M}=M_s(1-q)$. Let the matrix $\B{S} \in \mathbb{R}^{M \times N}$ consist of columns of the signatures from all neighbors of node $Z_0$, observable during the $M$ off-slots of node $Z_0$, and then normalized by $\sqrt{M_sq(1-q)}$ so that the expected value of the Euclidean norm of each column in $\B{S}$ is equal to $1$. \LZ{The number of rows in $\B{S}$ for each node varies depending on the number of off-slots in that node's signature, where the standard deviation of the fluctuation in percentage is about $1/\sqrt{M_s}$, which is quite small in cases of practical interest.}
Based on~\eqref{eq:PhyModel2}, the $M$-vector observed through all off-slots of node $Z_0$ can be expressed as
\begin{equation} \label{eq:CSModel}
  \B{Y}=\sqrt{\gamma_s}\B{S}\B{X}+\B{W}
\end{equation}
where
\begin{align}
\gamma_s=\gamma M_sq(1-q)/\sigma^2,
\end{align}
$\B{W}$ consists of circularly symmetric complex Gaussian entries with unit variance,
and $\B{X}$ is an $N$-vector indicating which $K$ signatures are selected to form the sum in~\eqref{eq:PhyModel2} as well as the signal strength for each neighbor. Precisely,
\begin{align*}
X_{(j-1)2^l+i} =\LZ{U_{0j}}\ind{d_j=i}
\end{align*}
for $1 \leq j \leq K$ and $1 \leq i \leq 2^l$. For example, for $K=3$ neighbors with $l=2$ bits of information each, where the messages are $d_1=3$, $d_2=2$, and $d_3=1$,
we have
\begin{equation}
\B{X}=[0\;\;0\;\;U_{01}\;\;0\;\;0\;\;U_{02}\;\;0\;\;0\;\;U_{03}\;\;0\;\;0\;\;0]^\LZH{\top}
\end{equation}
\LZH{where $[\cdot]^\top$ represents the transpose of a vector.} The sparsity of $\B{X}$ is exactly $2^{-l}$, which is typically very small. The average system load is defined as
\begin{align}
\beta=\frac{\expect{N}}{M_s(1-q)}=\frac{2^lc}{M_s(1-q)}
\end{align}
\LZ{where $c$ is the average number of neighbors defined in~\eqref{eq:c}.}

\DG{
An illustration of the RODD signaling scheme is given in Fig.~\ref{fig:signals}(a)--(b).  Fig.~\ref{fig:signals}(a) plots the transmitted signal of a node in three consecutive frame intervals, separated by dash-dotted lines.  Each frame carries one signature representing a message, which consists of mostly off-slots and a smaller number of on-slots, where the signal takes the values of $\pm1$.
Fig.~\ref{fig:signals}(b) plots the received signal of the same node, which is the superposition of all neighboring nodes' transmissions subject to fading, corrupted by noise and interference from non-neighbors.
The received signal is erased whenever the node transmits (hence the blank segments in the waveform).
}

Given the received signal, the decoding problem node $Z_0$ faces is to identify, out of a total of $N=2^lK$ signatures from all its neighbors, which $K$ signatures were selected. This requires every node to know the codebooks of all neighbors. One solution is to let the codebook of each node be generated using a pseudo-random number generator using its network interface address (NIA) as the seed, so that it suffices to acquire all neighbors' NIAs. This, in turn, is a neighbor discovery problem, which has been studied in~\cite{BorEph07AdhocNet, LuoGuo08Allerton, zhang2013neighbor}. The discovery scheme proposed in~\cite{LuoGuo08Allerton, zhang2013neighbor} uses similar on-off signaling and solves a compressed sensing problem.

\section{Sparse Recovery (Decoding) via Message Passing}
\label{sec:DecMB}
The problem of recovering the support of the sparse input $\B{X}$ based on the observation $\B{Y}$ has been studied in the compressed sensing literature. In this section, we develop an iterative message-passing algorithm based on belief propagation. The reasons for the choice include: 1) It is one of the most competitive decoding schemes in terms of error performance; and 2) the complexity is only linear in the dimensionality of the signal to be estimated.
BP belongs to a general class of message-passing
algorithms for statistical inference on graphical models, which
has demonstrated empirical
success in many applications including error-control codes, neural
networks, and multiuser detection in code-division multiple access
(CDMA) systems.

\subsection{The Factor Graph} \label{sec:BP}
\begin{figure}
\centering
\includegraphics[width=\columnwidth]{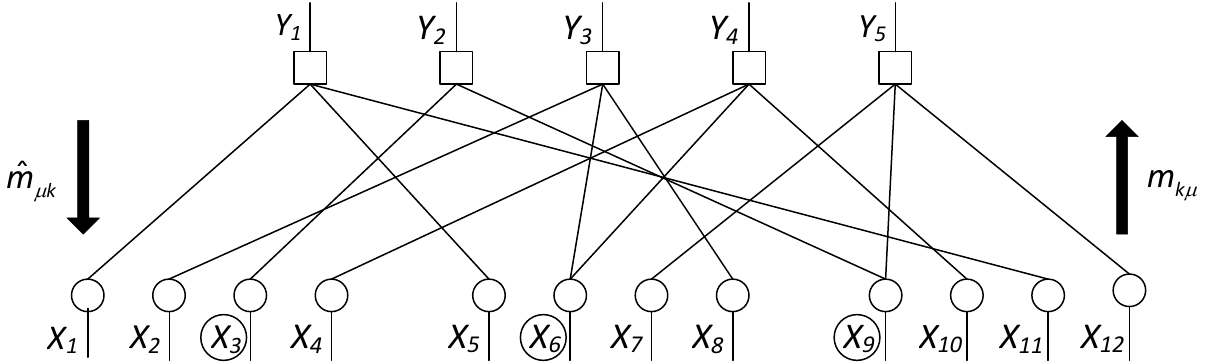}
\caption{The Forney-style factor graph of coded mutual broadcasting.
}
\label{fig:FG}
\end{figure}

In order to apply BP to mutual broadcasting, we construct a Forney-style bipartite factor graph to
represent the model~\eqref{eq:CSModel}. Here, we separate the real and imaginary parts in~\eqref{eq:CSModel} as
\begin{equation} \label{eq:CSModel_sep}
\B{Y}^{(1)}=\sqrt{\gamma_s}\B{S}\B{X}^{(1)}+\B{W}^{(1)},\,
\B{Y}^{(2)}=\sqrt{\gamma_s}\B{S}\B{X}^{(2)}+\B{W}^{(2)}
\end{equation}
where the superscripts $(1)$ and $(2)$ represent the real and imaginary parts respectively, $\B{W}^{(i)},i=1,2$ consists of i.i.d. Gaussian random variables with zero mean and variance $1/2$. The message-passing algorithm we shall develop based on~\eqref{eq:CSModel_sep} is not optimal, but such separation facilitates approximation and computation, which will be discussed in Section~\ref{sec:algo}. Since two parts in~\eqref{eq:CSModel_sep} share the same factor graph, we treat one of them and omit the superscripts:
\begin{equation} \label{eq:BPModel}
y_\mu=\sqrt{\gamma_s}\sum_{k=1}^{N}s_{\mu k}x_k+w_{\mu}
\end{equation}
where $\mu \in {1,2,\dots,M}$ and $k \in {1,2,\dots,N}$ index the
measurements and the input ``symbols,'' respectively.  For simplicity,
we ignore the dependence of the symbols $\{X_k\}$ for now, which shall
be addressed toward the end of this section.  Each $X_k$ then
corresponds to a symbol node and each $Y_{\mu}$ corresponds to a
measurement node, where the joint distribution of all $\{X_k\}$ and
$\{Y_{\mu}\}$ are decomposed into a product of $M+N$ factors, one
corresponding to each node.
For every $(\mu,k)$, symbol node $k$ and
measurement node $\mu$ are connected by an edge if $s_{\mu k}\neq 0$.
A simple example is shown in Fig.~\ref{fig:FG} for 5 measurements
and 3 neighbors each with 4 messages, i.e., $M=5$, $K=3$, and
$N=3\times4=12$.  The actual messages chosen by the three neighbors,
$d_1=3$, $d_2=2$ and $d_3=1$, correspond to the circled variables $X_3$, $X_6$ and $X_9$, respectively.

\subsection{The Message-Passing Algorithm} \label{sec:algo}

In general, an iterative message-passing algorithm involves two steps
in each iteration, where a message (or belief, which shall be
distinguished from an information message) is first sent from each
symbol node to every measurement node it is connected to, and then a
new set of messages are computed and sent in the reverse direction,
and so forth.  The algorithm performs exact inference within finite
number of iterations if there are no loops in the graph (the graph
becomes a tree if it remains connected).
In general, the algorithm attains a good approximate solution for
loopy graphs as the one in the current problem.

For convenience, let $\partial \mu$ (resp. $\partial k$) denote the
subset of symbol nodes (resp. measurement nodes) connected directly to
measurement node $\mu$ (resp. symbol node $k$), called its
neighborhood.\footnote{This is to be distinguished from the notion of neighborhood in the wireless network defined in Section~\ref{sec:neighbor}}
Let $|\partial \mu|$ (resp. $|\partial k|$) represent the cardinality of the neighborhood of measurement node $\mu$ (resp. symbol node $k$).
Also, let $\partial \mu \backslash k$ denote the neighborhood of
measurement node $\mu$ excluding symbol node $k$ and let $\partial k
\backslash \mu$ be similarly defined.
%

\begin{algorithm}
\caption{Message-Passing Decoding Algorithm}
\label{BP Alg}
\begin{algorithmic}[1]
\STATE {\it Input:} $\B{S},\B{Y},\gamma_s,M_s,q$.
\STATE {\it Initialization:}
\STATE $z_{\mu k}^{0,i} \leftarrow y_\mu^i/(\sqrt{\gamma_s}s_{\mu k})$ for all $s_{\mu k} \neq 0$ and $i=1,2$.
\STATE Initialize $\tau^{0,i}$ to a large positive number for $i=1,2$.
\STATE {\it Main iterations:}
\FOR{$t=1$ to $T-1$}
    \FORALL{$\mu$, $k$ with $s_{\mu k} \neq 0$ and $i=1,2$}
        \STATE $m_{k\mu}^{t,i} \leftarrow \expect{X\,\bigg|\,Y=\frac{\sum_{\nu \in \partial k \backslash \mu}z_{\nu k}^{t-1,i}}{|\partial k|-1};\frac{\tau^{t-1,i}}{{|\partial k|-1}}}$\,.
        \STATE $(\sigma_{k\mu}^{t,i})^2 \leftarrow \var{X\,\bigg|\,Y=\frac{\sum_{\nu \in \partial k \backslash \mu}z_{\nu k}^{t-1,i}}{|\partial k|-1};\frac{\tau^{t-1,i}}{{|\partial k|-1}}}$\,.
        \STATE $z_{\mu k}^{t,i} \leftarrow \frac{1}{\sqrt{\gamma_s}s_{\mu k}} \left(y_\mu^i-\sqrt{\gamma_s}\sum_{j \in \partial \mu \backslash k}s_{\mu j}m_{j\mu}^{t,i}\right)$\,.
    \ENDFOR
    \STATE $\tau^{t,i} \leftarrow \frac{1}{\sum_\mu|\partial \mu|}\sum_\mu |\partial \mu|\sum_{j \in \partial \mu}(\sigma_{j\mu}^{t,i})^2+\frac{1}{2\gamma_s}M_sq(1-q)$ for $i=1,2$.
\ENDFOR
\STATE $m_k^i \leftarrow \expect{X\,\bigg|\,Y=\frac{\sum_{\nu \in \partial k}z_{\nu k}^{T-1,i}}{|\partial k|-1};\frac{\tau^{T-1,i}}{{|\partial k|-1}}}$ for all $k$, $i=1,2$.
\STATE {\it Output:}
$\hat{w}_k = \arg\max_{j=1,\dots,2^l} \big|m_{(k-1)2^l+j}^1+\sqrt{-1}\,m_{(k-1)2^l+j}^2\big|$, $k=1,\dots,K$.
\end{algorithmic}
\end{algorithm}

The message-passing algorithm, given as Algorithm~\ref{BP Alg}, decodes the information indexes $d_1,\dots,d_K$, and is ready for implementation. \LZ{It is based on the conventional belief propagation algorithm. Central limit theorem and some other approximation techniques are used to
reduce the computational complexity, which will be discussed in detail shortly.}

 The superscripts $i=1,2$ in Algorithm~\ref{BP Alg} represent the real and imaginary parts, respectively. Here, $\expect{X\,\big|\,Y=y;\LZH{\xi}}$ and $\var{X\,\big|\,Y=y;\LZH{\xi}}$ represents the conditional mean and variance of the input given the Gaussian channel output $Y=X+W$ with $W \sim \mathcal{N}(0,\LZH{\xi})$ is equal to $y$. Mathematically, assume $X$ has cumulative distribution function $P_X(x)$, then for $i=1,2,\dots$,
\begin{equation} \label{eq:condExp}
\expect{X^i\,\big|\,Y=y;\sigma^2}=\frac{\int x^i e^{-\frac{(y-x)^2}{2\LZH{\xi}}}\diff P_X(x)} {\int e^{-\frac{(y-x)^2}{2\LZH{\xi}}}\diff P_X(x)}
\end{equation}
and
\begin{align}
  \begin{split}
  \var{X\,\big|\,Y=y;\LZH{\xi}}
  &= \expect{X^2\,\big|\,Y=y;\LZH{\xi}} \\ 
  &\qquad -\left(\expect{X\,\big|\,Y=y;\LZH{\xi}}\right)^2
  \end{split}
\end{align}
where $\int \cdot\,\diff P_X(x)$ 
denotes the Riemann-Stieltjes integral.

In the following, we derive Algorithm~\ref{BP Alg} starting from~\eqref{eq:BPModel} which is valid for both real and imaginary parts in~\eqref{eq:CSModel_sep}. It is a simplification of the original iterative BP algorithm, which iteratively computes the marginal {\em a posteriori} distribution of all symbols given the measurements, assuming that the graph is free of cycles. For each $k\in \partial \mu$ (hence $\mu \in\partial k$), let $\big\{V^t_{k\mu}(x)\big\}$ represent the message from symbol node $k$ to measurement node $\mu$ at the $t$-th iteration and $\big\{U^t_{\mu k}(x)\big\}$ represent the message in the reverse direction. Each message is basically the belief (in terms of a probability density or mass function) the algorithm has accumulated about the corresponding symbol based on the measurements on the subgraph traversed so far, assuming it is a tree. Let $p_X(x)$ denote the {\em a priori} probability density function of $X$. In the $t$-th iteration, \LZ{as in the belief propagation algorithm~\cite{TanRas06ISIT, GuoWan08JSAC},} \LZnew{$V_{k\mu}^t(x)$ is computed by combining the prior information $p_X(x)$ and messages from measurement nodes in the previous iteration $U_{\nu k}^{t-1}(x)$, and $U_{\mu k}^t(x)$ is calculated based on the Gaussian channel model and the messages from symbol nodes \LZH{in the current iteration $V_{k\mu}^t(x)$}. 
Therefore, we have}
\begin{subequations} \label{eq:vv}
\begin{align}
  V_{k\mu}^t(x) &\propto p_X(x)\prod_{\nu \in \partial k \backslash
    \mu}U_{\nu k}^{t-1}(x) \label{eq:V}
\end{align}
for all $(k,\mu)$ with $s_{\mu k}\ne0$, and then
\begin{align}
  \begin{split}
  U_{\mu k}^t(x) &\propto \int_{(x_j)_{\partial \mu \backslash k}}
  \exp\bigg[-\big(y_{\mu}-\sqrt{\gamma_s}s_{\mu k}x \\
  -&\sqrt{\gamma_s}\sum_{j \in \partial \mu \backslash k}s_{\mu j}x_j
  \big)^2 \bigg] \LZnew{\prod_{j \in \partial \mu \backslash k} \left(V_{j\mu}^t(x_j)\diff x_j\right)} \label{eq:U}
  \end{split}
\end{align}
\end{subequations}
where $\int_{(x_j)_{\partial \mu \backslash k}}$ denotes integral over all
$x_j$ with $j \in \partial \mu \backslash k$, and $V(x) \propto
u(x)$ means that $V(x)$ is proportional to $u(x)$ with proper
normalization such that $\int_{-\infty}^{\infty} V(x)\diff x=1$. In case $X$ is a discrete random variable, the integral shall be replaced by a sum over the alphabet of $X$. In this problem, $X$ follows a mixture of discrete and continuous distributions, so the expectation can be decomposed as an integral and a sum.

The complexity of computing the integral in~\eqref{eq:U} is exponential in
$|\partial \mu|=\mathcal{O}(qN)$, which is in general infeasible for
the problem at hand.
However, as $qN\gg1$, the computation carried out at each measurement
node admits a good approximation by using the central limit theorem.  A
similar technique has been used in the CDMA detection problem, for
fully-connected bipartite graph in~\cite{Kab03JPAMG, TanOka05IT,
  TanRas06ISIT}, and for a graph with large node degrees
in~\cite{GuoWan08JSAC}.

To streamline~\eqref{eq:V} and~\eqref{eq:U}, we introduce $m_{k\mu}^t$ and $(\sigma_{k\mu}^t)^2$ for all $(\mu,k)$ pairs with $s_{\mu k}\ne0$ to represent the mean and variance of a random variable with distribution $V_{k\mu}^t(x)$.
Using Gaussian approximation, one can reduce the message-passing
algorithm to iteratively computing the following messages with the initial conditions that $z_{\mu k}^0=y_{\mu}/\left(\sqrt{\gamma_s}s_{\mu k}\right)$ and $\tau^0$ is a large positive number:
\begin{subequations} \label{eq:iter}
\begin{align}
m_{k\mu}^t &= \expect{X\,\bigg|\,Y=\frac{\sum_{\nu \in \partial
k \backslash \mu}z_{\nu k}^{t-1}}{|\partial k|-1};\frac{\tau^{t-1}}{{|\partial k|-1}}} \label{eq:condMean} \\
(\sigma_{k\mu}^t)^2 &= \var{X\,\bigg|\,Y=\frac{\sum_{\nu \in \partial
k \backslash \mu}z_{\nu k}^{t-1}}{|\partial k|-1};\frac{\tau^{t-1}}{{|\partial k|-1}}} \label{eq:condVar} \\
z_{\mu k}^t &= \frac{1}{\sqrt{\gamma_s}s_{\mu k}} \left(y_\mu-\sqrt{\gamma_s}\sum_{j \in \partial \mu \backslash k}s_{\mu j}m_{j\mu}^t\right) \label{eq:z} \\
\tau^t &= \frac{1}{\sum_\mu|\partial \mu|}\sum_\mu |\partial \mu|\sum_{j \in \partial \mu}(\sigma_{j\mu}^t)^2+\frac{1}{2\gamma_s}M_sq(1-q) \label{eq:tau}
\end{align}
\end{subequations}
where~\eqref{eq:condMean} and~\eqref{eq:condVar} calculate the conditional expectation and variance, respectively. The detailed derivation is relegated to Appendix~\ref{a:bp}. At the $T$-th iteration, the approximated posterior mean of $x_k$ can be expressed as
\begin{equation} \label{eq:apm}
m_k = \expect{X\,\bigg|\,Y=\frac{\sum_{\nu \in \partial
k}z_{\nu k}^{T-1}}{|\partial k|-1};\frac{\tau^{T-1}}{{|\partial k|-1}}} \,.
\end{equation}

It is time consuming to compute~\eqref{eq:condMean} and~\eqref{eq:condVar} for all $(\mu,k)$ pairs with $s_{\mu k}\ne 0$, especially in the case of large matrix $\B{S}$. We can use the following two approximation techniques to further reduce the computational complexity. First, $|\partial k|$ in~\eqref{eq:condMean},~\eqref{eq:condVar} and~\eqref{eq:apm} is replaced by its mean value $M_sq(1-q)$. Second, we use interpolation and extrapolation to further reduce the computation complexity of~\eqref{eq:condMean},~\eqref{eq:condVar} and~\eqref{eq:apm}. Specifically, in each iteration $t$, we only compute the conditional mean and variance for some chosen $y$'s, i.e., we choose $y^t_1<y^t_2<\dots<y^t_n$ which is a partition of an interval depending on $\tau^{t-1}$, compute
\begin{align}
a_j^t &= \expect{X\,\bigg|\,Y=y_j^t;\frac{\tau^{t-1}}{M_sq(1-q)-1}}
\\
b_j^t &= \var{X\,\bigg|\,Y=y_j^t;\frac{\tau^{t-1}}{M_sq(1-q)-1}}
\end{align}
for $j=1,2,\dots,n$, and then use those values to calculate~\eqref{eq:condMean},~\eqref{eq:condVar} \LZnew{and~\eqref{eq:apm}} by interpolation or extrapolation. To be more precise, for any pair $\mu,k$ with $s_{\mu k}\neq 0$, suppose $y_j^t$ and $y_{j+1}^t$ are chosen to be the closest to
\begin{align}
  \label{eq:1}
  y=\frac{\sum_{\nu \in \partial k \backslash \mu}z_{\nu k}^{t-1,i}}{|\partial k|-1},
\end{align}
then $m_{k\mu}^t$ \LZnew{in~\eqref{eq:condMean}} and $(\sigma_{k\mu}^t)^2$ \LZnew{in~\eqref{eq:condVar}} can be approximated by
\begin{align}
m_{k\mu}^t &= a_j^t+\frac{y-y_j^t}{y_{j+1}^t-y_j^t}(a_{j+1}^t-a_j^t)
\\
(\sigma_{k\mu}^t)^2 &= b_j^t+\frac{y-y_j^t}{y_{j+1}^t-y_j^t}(b_{j+1}^t-b_j^t).
\end{align}
Similarly, \LZnew{$m_k$ in~\eqref{eq:apm} can be approximately calculated.}

We now revisit the assumption that $\B{X}$ has independent elements.
In fact, $\B{X}$ consists of $K$ sub-vectors of length $2^l$, where
the entries of each sub-vector are all zero except for one position
corresponding to the transmitted message.
After obtaining the approximated posterior mean $\widetilde{m}_k$ by incorporating both real and imaginary parts calculated from~\eqref{eq:apm}, Algorithm~\ref{BP Alg} outputs the position of the element with the largest magnitude in each of the $K$ sub-vectors of $[\widetilde{m}_1,\dots,\widetilde{m}_N]$.
In fact the factor graph Fig.~\ref{fig:FG} can be modified to
include $K$ additional nodes, each of which puts a constraint on one
sub-vector.  Slight improvement over Algorithm~\ref{BP Alg} 
may be
obtained by carrying out message passing on the modified graph.

\LZ{
The performance of Algorithm 1 has been analyzed in~\cite[Section 4.4.3]{Zhang12PhD} in the so-called {\em large-system limit}, where the frame length and the number of messages a node sends both tend to infinity with a fixed ratio.  The evolution of the error rate achieved by Algorithm 1 with different number of iterations is asymptotically characterized by a fixed-point equation.  The analysis uses techniques developed for statistical inference through noisy large linear systems.  It is not the focus of this paper and thus is omitted.

The number of iterations, $T$, needed to achieve good performance in practice is typically not large.  In the simulation results given in Section VI, we use $T=16$ iterations.
}

\section{Random Access Schemes} \label{sec:Compare}

In this section we describe two random access schemes, namely slotted ALOHA and CSMA, and provide lower bounds on the message error probability.  The results will be used in Section~\ref{sec:nr} to compare with the performance of RODD.

\LZ{Suppose node transmissions are synchronized. The nominal SNR in each slot is the same as in~\eqref{eq:PhyModel}, also denoted by $\gamma$.} \DG{The channel model, network model and propagation model (including Rayleigh fading) are as introduced in Section~\ref{sec:Model}.}

Let $L$ denote the total number of bits encoded into a frame, which includes an $l$-bit message and a few additional bits which identify the sender. This is in contrast to broadcasting via
compressed sensing, where the signature itself identifies the sender (and carries the message). Each broadcasting period consists of a number of frames to allow for retransmissions. \LZ{A message is assumed to be decoded correctly if the signal-to-interference-plus-noise ratio (SINR) in the corresponding frame transmission exceeds 
a threshold $\delta$ (multi-packet reception is possible only if $\delta<1$). Over the additive white noise channel with SINR $\delta$, in order to send $L$ bits reliably through the channel, the number of symbols in a frame must exceed $L/\log_2(1+\delta)$. Therefore, the number of frames in a period of $M_r$ symbol intervals should satisfy
\begin{equation} \label{eq:UBframes}
N_r \leq M_r\log_2(1+\delta)/L.
\end{equation}
}

Without loss of generality, we still consider the typical node $Z_0$ at the origin. An error event is defined as that node $Z_0$ cannot correctly recover the message from one specific neighbor \LZ{after a period of $M_r$ symbol intervals.}  The corresponding error probabilities achieved by slotted ALOHA and CSMA are denote by $\Probt{a}{e}$ and $\Probt{c}{e}$, respectively. \LZ{For ease of discusssion, we
allow the total number of symbol intervals, $M_r$, to be any positive integer, which may not be a multiple of the frame length.  This results in underestimated number of intervals needed by the random access scheme to attain the desired performance. }

\subsection{Slotted ALOHA} \label{sec:aloha}
In slotted ALOHA, suppose each node chooses independently with the same probability $p$ to transmit in every frame interval.
\DG{Fig.~\ref{fig:signals}(c) illustrates signals transmitted by a typical node over 20 frame intervals, separated by dash dotted lines.  In each of the 6 active frame intervals, capacity-achieving Gaussian signaling is used.  The node listens to the channel over the remaining frame intervals to receive the signal shown in Fig.~\ref{fig:signals}(d).  The received signals during the node's own transmitting frames are erased.  During some frame intervals, the received signals appear to be strong, which implies that one or more neighbors have transmitted.  During some other frame intervals, the received signals appear to be weak, which consist of only noise and interference from non-neighbors.
}

Let $Z$ denote one specific neighbor of node $Z_0$ and $G$ denote the fading coefficient between them. Suppose the mark of $Z$ is denoted by $\mathcal{G}$. Given that $(Z,\mathcal{G}) \in \hat{\Phi}$ where $\hat{\Phi}$ is given by~\eqref{eq:Phihat}, denote $\hat{\Phi}_1=\hat{\Phi} \backslash \{(Z,\mathcal{G})\}$, which is also a marked p.p.p. with intensity $\lambda$. For a given realization of $(Z,G)$ and $\hat{\Phi}_1$, define $\Probt{a}{s}(Z,G,\hat{\Phi}_1)$ as the probability that the received SINR from $Z$ to $Z_0$ exceeds 
the threshold $\delta$ conditioning on that $Z$ transmits in a given frame. In any given frame, the probability of the event that $Z$ transmits, $Z_0$ listens, and the transmission is successful is thus $p(1-p)\Probt{a}{s}(Z,G,\hat{\Phi}_1)$. Therefore, the probability that the message from $Z$ has not been successfully received by $Z_0$ after $\LZ{N_r}$ consecutive frame intervals can be expressed as
\begin{equation} \label{eq:Perr}
\Probt{a}{e} = \expect{\left(1-p(1-p)\Probt{a}{s}(Z,G,\hat{\Phi}_1)\right)^{\LZ{N_r}}}
\end{equation}
where the expectation is over the joint distribution $(Z,G,\hat{\Phi}_1)$. Due to the convexity of function $\left(\max\{0,1-z\}\right)^n$, $z\geq0,n\in\{1,2,\dots\}$, $\Probt{a}{e}$ in~\eqref{eq:Perr} can be lower bounded as
\begin{equation} \label{eq:pe_lb}
\Probt{a}{e} \geq \left( \max\left\{0,1-p(1-p)\expect{\Probt{a}{s}(Z,G,\hat{\Phi}_1)} \right\}\right)^{\LZ{N_r}} .
\end{equation}

In Appendix~\ref{a:aloha}, the expectation of $\Probt{a}{s}(Z,G,\hat{\Phi}_1)$ is \LZnew{upper bounded}
using the known Laplace transform of the distribution of the interference~\cite{BacBla09FT}.
\LZ{For a period of $M_r$ symbol intervals}, the lower bound on $\Probt{a}{e}$ is presented in the following result.

\begin{proposition} \label{prop:Pelb}
Consider an arbitrary neighbor $Z$ of node $Z_0$. The probability that $Z_0$ cannot successfully receive the message from $Z$ after \LZ{a period of $M_r$ symbol intervals} is lower bounded as follows:
\begin{align} \label{eq:aloha}
\Probt{a}{e} \geq& \Bigg(\max\Bigg\{0,1-\frac{1}{\pi} p(1-p) \left(\frac{\theta}{\delta}\right)^b\sin\left(\frac{b\pi}{2}\right)\Gamma(1-b) \nonumber \\
&\int_{-\infty}^\infty |\omega|^{b-1} \exp\left\{-\lambda p\frac{b\pi^2}{\sin(b\pi)}(\iota\omega)^b-\iota\frac{\omega}{\gamma}\right\} \diff\omega \Bigg\}\Bigg)^{\LZ{n_r}}
\end{align}
where $\iota=\sqrt{-1}$, $b=2/\alpha$ and $\LZ{n_r}=\LZ{M_r}\log_2{(1+\delta)}/L$.
\end{proposition}

Although~\eqref{eq:aloha} appears to be complicated, computing it only involves a straightforward single-variable integral (the outcome of the integral is in fact real-valued).

In the slotted ALOHA scheme, despite repeated transmissions, a given link may still fail to deliver the message due to the half-duplex constraint (the receiver happens to transmit during the same frame) and consistently weak received SINR due to random interference from other links.

\subsection{CSMA} \label{sec:csma}
As an improvement over ALOHA, CSMA lets nodes use a brief contention
period to negotiate a schedule in such a way that nodes in a small
neighborhood do not transmit data simultaneously. We analyze the performance of CSMA by using the Mat\'{e}rn hard core model~\cite{BacBla09FT}. To be specific, consider the following generic scheme: Each node senses the channel continuously; if the channel is busy, the node remains silent and disables its timer; as soon as the channel becomes available, the node starts its timer with a random offset, and waits till the timer expires to transmit its frame. Clearly, the node whose timer expires first in its neighborhood captures the channel and transmits its frame.

Mathematically, let $\{T_i\}$ be i.i.d.\ random variables with uniform distribution on $[0,1]$, which represent the timer offsets for all nodes $\{Z_i\}$ in $\Phi$, respectively. \LZnew{Node $Z_i$ will transmit its frame if and only if $T_i<T_j$ for all $X_j\in\mathcal{N}(Z_i)$.}

\LZnew{By viewing $T_i$ as a mark of node $Z_i$, we redefine $\tilde{\Phi}$ and $\hat{\Phi}$ as $\tilde{\Phi}=\{(Z_i,\mathcal{G}_i,T_i)\}_i$ and $\hat{\Phi}=\tilde{\Phi}\backslash (Z_0,\mathcal{G}_0,T_0)$, respectively}. Let $Z$ be one specific neighbor of $Z_0$ \LZnew{and $T$ denote its time offset}. Define $G$ and $\mathcal{G}$ as in Section~\ref{sec:aloha}. Given that $(Z,\mathcal{G}\LZnew{,T}) \in \hat{\Phi}$, denote $\hat{\Phi}_1=\hat{\Phi} \backslash \{(Z,\mathcal{G}\LZnew{,T})\}$, \LZnew{which is still a marked p.p.p. with intensity $\lambda$}. For a given realization of $(Z,G\LZnew{,T})$ and $\hat{\Phi}_1$, define $\Probt{c}{s}(Z,G\LZnew{,T},\hat{\Phi}_1)$ as the probability that node $Z$ transmits its frame and the received SINR from $Z$ to $Z_0$ exceeds 
the threshold $\delta$. Therefore, the probability that the message from $Z$ has not been successfully received after $\LZ{N_r}$ consecutive frame intervals can be expressed as
\begin{align}
\Probt{c}{e} &= \expect{\left(1-\Probt{c}{s}(Z,G\LZnew{,T},\hat{\Phi}_1)\right) ^{\LZ{N_r}}} \nonumber \\
&\geq \left( \max\left\{0,1-\expect{\Probt{c}{s}(Z,G\LZnew{,T},\hat{\Phi}_1)} \right\}\right)^{\LZ{N_r}} \label{eq:pe_lb2}
\end{align}
where the expectation is over the joint distribution of $(Z,G\LZnew{,T},\hat{\Phi}_1)$, and~\eqref{eq:pe_lb2} is due to the convexity of function $\left(\max\{0,1-z\}\right)^n$, $z\geq0,n \in\{1,2\dots\}$.

\LZ{For a period of $M_r$ symbol intervals}, the lower bound on error probability $\Probt{c}{e}$ is given by the following result, which is proved in Appendix~\ref{a:csma}.

\begin{proposition} \label{prop:Pelb2}
Consider an arbitrary neighbor $Z$ of node $Z_0$. The probability that $Z_0$ cannot successfully receive the message from $Z$ after \LZ{a period of $M_r$ symbol intervals} is lower bounded as follows:
\begin{equation} \label{eq:csma}
\Probt{c}{e} \geq \left(\max\left\{0,1-\frac{1}{c^2} \left(\frac{\theta\gamma}{\delta}\right)^{\frac{2}{\alpha}} \left(e^{-c}+c-1\right) \right\}\right)^{\LZ{n_r}}
\end{equation}
where $c$ is defined in~\eqref{eq:c} and $\LZ{n_r}=\LZ{M_r}\log_2{(1+\delta)}/L$.
\end{proposition}

In contrast to slotted ALOHA, frame loss due to the half-duplex constraint is eliminated through contention. However, a given link may still fail to deliver the message after repeated transmissions because the received SINR were consistently weak due to random interference outside the neighborhood.

\section{Numerical Results}\label{sec:nr}
In order for a fair comparison, we assume the same power constraint
for both the 
compressed sensing
scheme and random access schemes, i.e., the average transmit power in each active slot (in which the node transmits energy) is the same.
We choose the same transmission probability in each slot for
\DGnew{the compressed sensing}
and slotted ALOHA \DGnew{schemes}, i.e., $q=p=1/(c+1)$. Also, the transmission probability in each slot for CSMA is $(1-e^{-c})/c$ (see Appendix~\LZ{\ref{a:csma}}), which is close to $1/(c+1)$ when $c$ is large. The three schemes consume approximately the same amount of average power over any period of time.

Without loss of generality, let one unit of distance be $1$ meter. Consider a wireless network of $1000$ nodes uniformly distributed in a square with side length of $500$~meters. The nodes form a Poisson point process in the square conditioned on the node population. Suppose the path-loss exponent $\alpha=4$. The threshold of channel gain to define neighborhood is set to $\theta=10^{-6}$. It means that if the SNR from a node one meter away is 60~dB, then the SNR attenuates to $0$~dB (the neighborhood boundary) at $10^{6/\alpha} \approx 31$~meters \LZ{due to path loss only. As both path loss and fading are considered, a node near the center of the square (without boundary effect) has on average $c\approx 11$ neighbors according to~\eqref{eq:c}.}

\begin{figure}[t]
\centering
\includegraphics[width=\columnwidth]{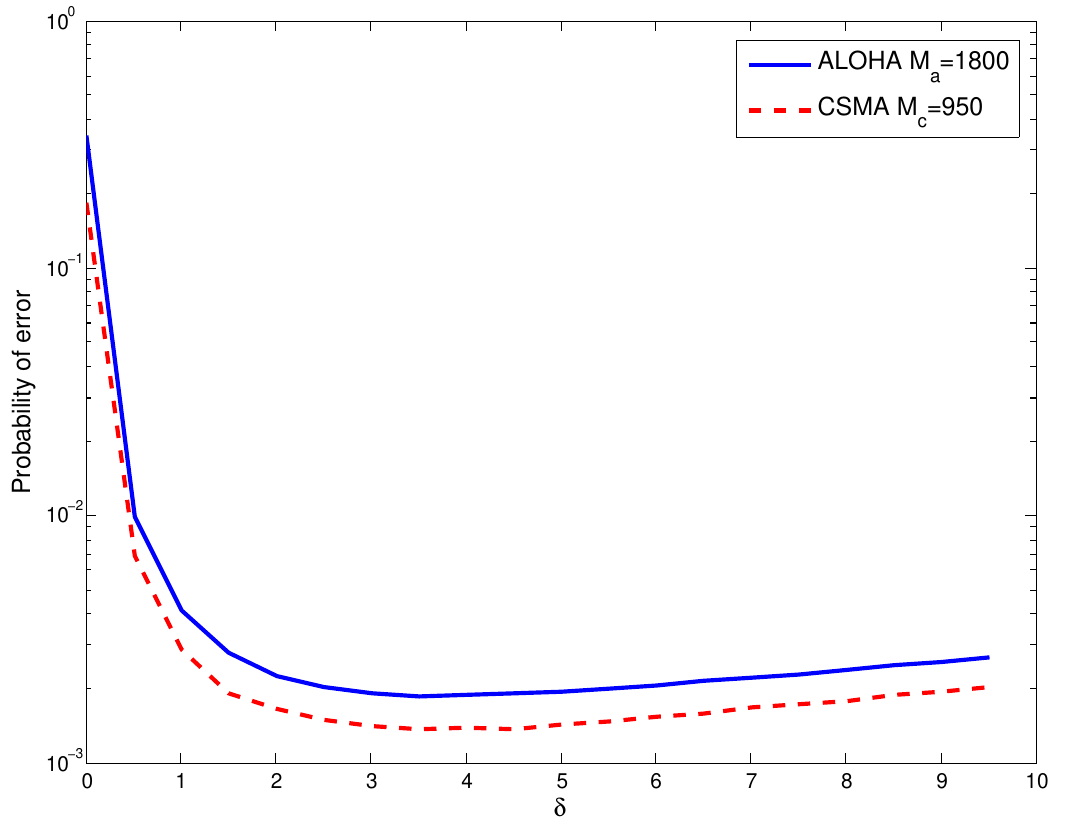}
\caption{\LZH{Lower} 
bounds for error probability in slotted-ALOHA and CSMA for different threshold $\delta$ in the case of $l=10$.} \label{fig:T}
\end{figure}

We consider two cases for the length of broadcasting, with $l=5$ and $10$~bits, respectively. In random access schemes, a packet of $L$~bits consists of $l$-bit message and $\ceil{\log_2c}$ additional bits to identify the sender. Fig.~\ref{fig:T} shows that $\LZ{\delta}=3.5$ minimizes the lower bounds for $\Probt{a}{e}$ in~\eqref{eq:aloha} and $\Probt{c}{e}$ in~\eqref{eq:csma} in the case of $l=10$.

\begin{figure}[t]
\centering
\includegraphics[width=\columnwidth]{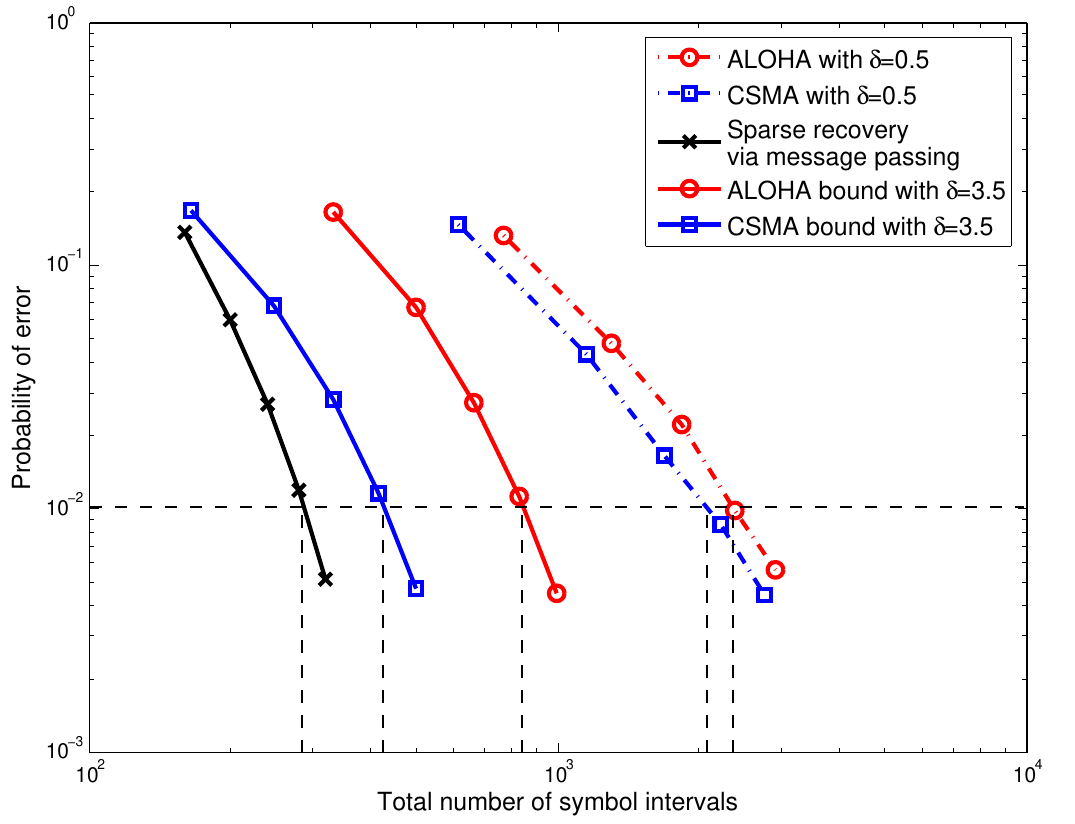}
\caption{Performance comparison between
\DGnew{the compressed sensing}
 and random access \DGnew{schemes}. Each node transmits a $5$-bit message.} \label{fig:L5}
\end{figure}

The metric for performance comparison is the probability for one node to miss one specific neighbor, averaged over all pairs of neighboring nodes in the network. Suppose the transmit SNR of each node is $\gamma=60$~dB. First consider one realization of the network where each node has $c\approx 11$ neighbors on average and $l=5$ bits to broadcast, so that on average $cl\approx 55$ bits are to be collected by each node. In slotted ALOHA, at least $4$ additional bits are needed to identify a sender out of $1000$ nodes, so we let $L=9$. In Fig.~\ref{fig:L5}, the error performance of slotted ALOHA and CSMA for $\delta=0.5$ is compared with that of the
compressed sensing
scheme with the message-passing algorithm. The simulation result shows the
compressed sensing
scheme significantly outperforms slotted ALOHA and CSMA, even compared with the minimum of the lower bounds computed from~\eqref{eq:aloha} and~\eqref{eq:csma} for $\delta=3.5$. For example, to achieve $1\%$ error rate, the
compressed sensing
scheme takes fewer than $300$ symbols. Slotted ALOHA and CSMA take no less than $800$ and $400$ symbols according to the bounds in~\eqref{eq:aloha} and~\eqref{eq:csma} \LZ{for $\delta=3.5$}, respectively. In fact, slotted ALOHA and CSMA with threshold $\delta=0.5$ take more than $2000$ symbols. Similar comparison is observed for several other SINR thresholds $\delta$ around $0.5$ and the performance of ALOHA and CSMA are not good for $\delta\geq 1$ because the messages from weaker neighbors may never be successfully delivered. Some additional
supporting numerical evidence is, however, omitted due to
space limitations.

\begin{figure}
\centering
\includegraphics[width=\columnwidth]{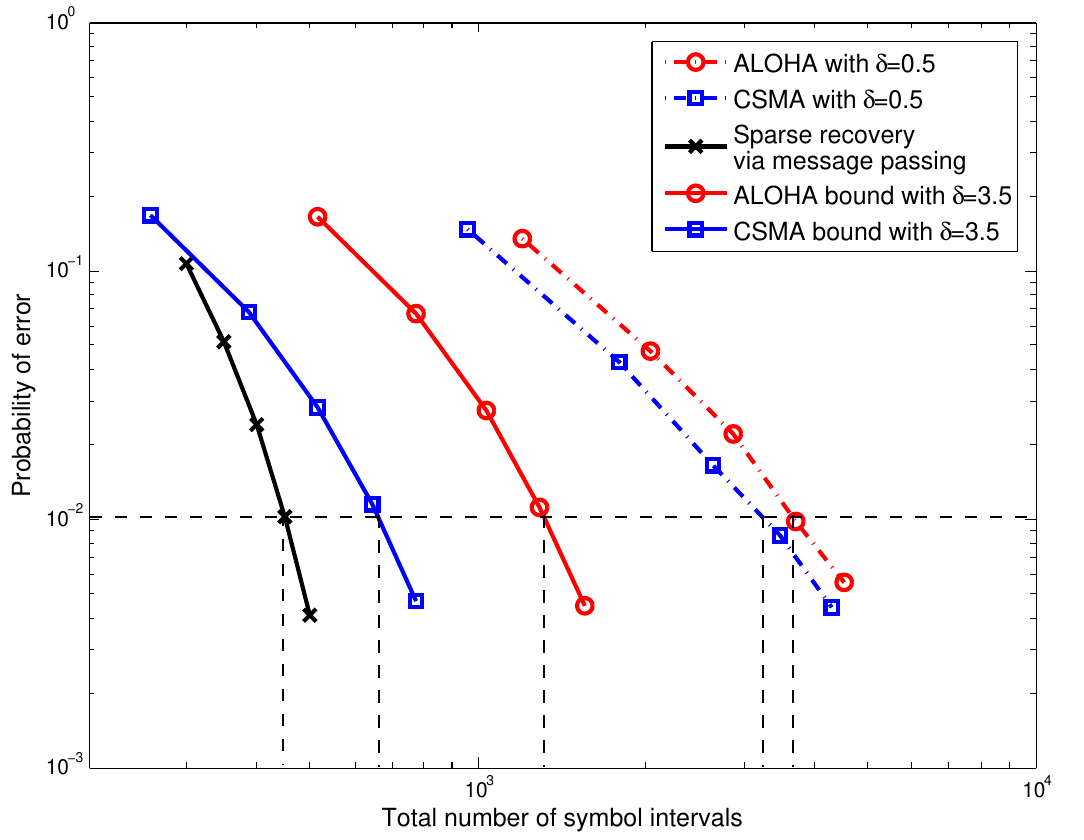}
\caption{Performance comparison between
\DGnew{the compressed sensing}
 and random access \DGnew{schemes}. Each node transmits a $10$-bit message.}
\label{fig:L10}
\end{figure}

\begin{figure}
\centering
\includegraphics[width=\columnwidth]{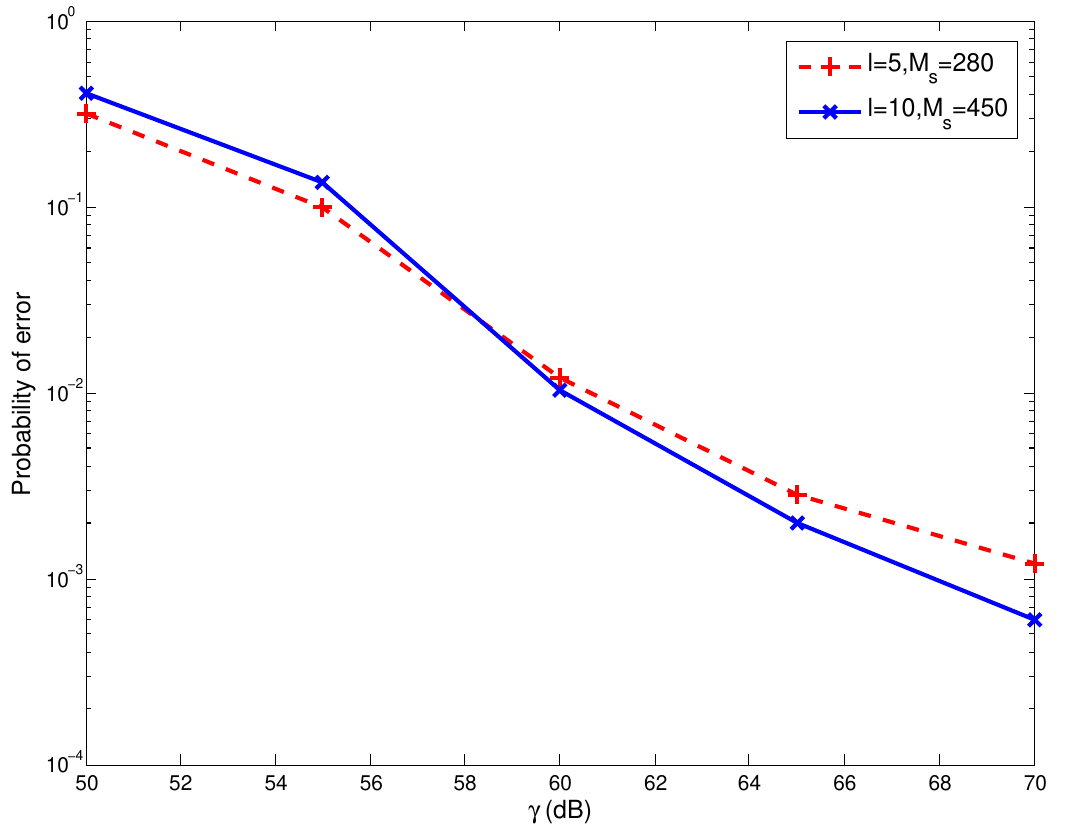}
\caption{Performance of
\DGnew{the compressed sensing}
scheme
with
different nominal SNRs ($\gamma$).}
\label{fig:50-70dB}
\end{figure}

\begin{figure}
\centering
\includegraphics[width=\columnwidth]{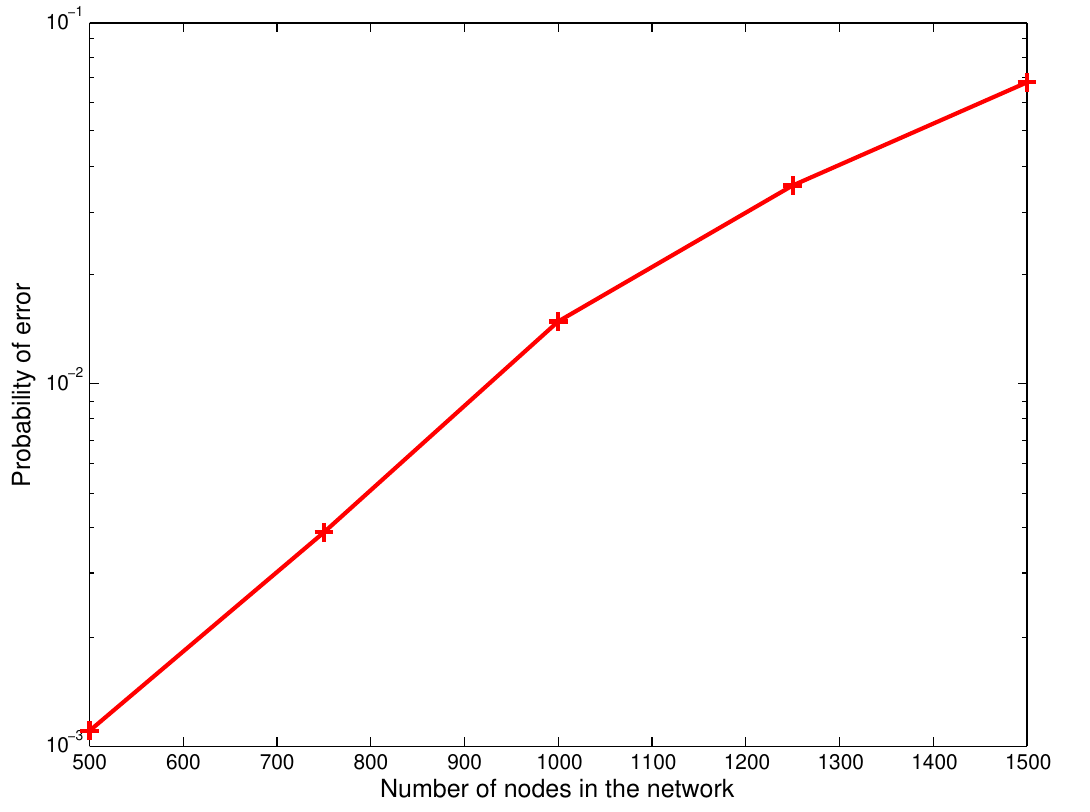}
\caption{Performance of
\DGnew{the compressed sensing}
 scheme
with different node densities.}
\label{fig:Pe(density)}
\end{figure}

Fig.~\ref{fig:L10} repeats the experiment of Fig.~\ref{fig:L5} with $10$-bit messages. The
compressed sensing
scheme has significant gain compared with slotted ALOHA and CSMA. For example, to achieve the error rate of $1\%$, the
compressed sensing
scheme takes about $450$ symbols, whereas slotted ALOHA and CSMA take at least $1000$ and $650$ symbols, respectively.

In Fig.~\ref{fig:50-70dB}, we simulate the same network with different nominal SNRs, i.e., $\gamma$ varies from $50$~dB to $70$~dB. In the case that each node transmits a $5$-bit message, the frame length is chosen to be $280$ symbols. It can be seen from the figure that the probability of error decreases with the increase of SNR. The performance is similar when
each node transmits a $10$-bit message and the frame consists of $450$ symbols.

Fig.~\ref{fig:Pe(density)} plots the probability of error achieved by our
compressed sensing
scheme as a function of the number of nodes on the 500$\times$500
$m^2$ square where all other parameters are held constant.  In
particular, the nominal SNR $\gamma=60$~dB.  The frame is of 280
symbol intervals and each message consists of 5 bits.  As the number
of nodes increases from 500 to 1500, the average number of neighbors a node has increases from about 5 to about 16, where the frame error rate increases gracefully from about 0.1\% to 7\%.


\section{Discussion and Concluding Remarks}
\label{sec:Conclude}

\DGnew{This study consists of two integrated components: One is the compressed sensing scheme for nodes to simultaneously broadcast to one-hop neighbors; the other is the on-off signaling that enables virtual full-duplex communication.
Importantly, if alternative techniques are used to enable full duplex, the compressed sensing scheme applies equally well, where the codewords need not be on-off.}

In the following, we provide further discussion of the proposed technology for wireless broadcasting, especially the advantages of RODD over related well known schemes.

\subsection{RODD vs.~CDMA}

\DGnew{In both RODD and direct sequence CDMA, nodes simultaneously transmit data bearing sequences.  Their similarity ends there.
In particular, their different timescales, access modes and duplex schemes set them apart:
1) A CDMA spreading sequence spans over one {\em symbol interval} to carry one symbol, while a RODD sequence or signal spans over one {\em frame interval} to represent one frame of data.
Typically, frames are the units of error control coding; that is, all symbols in a given frame form a codeword, whereas different frames are coded separately.
Thus the RODD sequence is one codeword, whereas in CDMA, one codeword spans over many spreading sequences (often repetitions of the same sequence);}
2) RODD is designed for many-to-many transmission, whereas CDMA is a many-to-one multiaccess scheme;
3) RODD signaling enables full-duplex communication at frame level, whereas CDMA is half duplex in the absence of self-interference cancellation.

\subsection{RODD vs.~TDMA}

Time-division multiple access (TDMA), which is suitable for many-to-one communication, is difficult to apply in many-to-many communication in a large network, where different nodes see different neighborhoods.
In the scenario considered in this paper, every node in a large network wishes to broadcast its message to all neighbors, and also wishes to receive messages from all neighbors.
To generate a time division schedule for all nodes to avoid collision in every neighborhood that is throughput optimal is an NP-hard problem.  Using ad hoc solutions leads to a highly conservative schedule with low throughoput, whereas using a more aggressive schedule causes collisions and require complicated scheduling of retransmissions.
Furthermore, a time-division schedule needs to be recomputed if a node moves in or out of a neighborhood, whereas the proposed RODD scheme is robust to network topology changes.

\subsection{RODD vs.~Random Access}

\DGnew{The on-off signaling of RODD resembles that of ALOHA at a much faster timescale.
There are, however, crucial differences.
As frames are usually the units of error control coding,
each frame (or a large on-slot) is coded separately in ALOHA or CSMA.}
In RODD, each frame consists of many (short) slots, so that we code over all the on-slots.
From an individual node's viewpoint, other nodes' transmissions are seen as interference.
If nodes use RODD signaling, the node \DGnew{experiences ergodic interference}, wheres if nodes use random access schemes, the node \DGnew{experiences nonergodic interference}.  The former channel is ergodic because the interference fluctuates at slot level, but statistically the channel remains the same in every frame.  The latter channel is nonergodic because the channel fluctuates at frame level, and appears very different in different frames.  The node may see no interference in some frames, but a lot of interference in other frames.  When the channel is nonergodic, the node cannot predict the interference level, hence the node does not know the best 
\DGnew{data} rate to \DGnew{transmit}. 
If the node 
\DGnew{transmits} at high rate, the frame may be lost when other nodes also transmit.  If the node 
\DGnew{transmits} at low rate, channel resources are wasted if no other nodes transmit at the same time.  This problem is entirely overcome by RODD signaling.  Since the channels of all frames look statistically the same for a given node, all frames can be coded at the same rate up to the capacity of the channel and be decoded 
\DGnew{reliably by the receiving node}. 

One might argue that it would be just as good to code over many frames when random access is used to also yield an ergodic channel.  The problem with this is that many frames have to be received before decoding them, hence the decoding delay would be exceedingly large.  For the same maximum decoding delay, random access achieves much lower rate compared to RODD.

\subsection{RODD vs.~Interference Cancellation}

State-of-the-art MAC protocols
\DGnew{are mostly designed based on the packet collision model for wireless networks,}
where if multiple nodes simultaneously transmit, their transmissions fail due to collision at the receiver.
\DGnew{In contrast, RODD signaling takes full advantage of the superposition nature of the wireless medium.
By coding over the entire frame of on- and off-slots and joint decoding of multiple users, signals in colliding slots are also fully utilized.}

Other recent works such as~\cite{HalAnd08Mobicom, GolKat08Sigcomm} break away from  the collision model.
 The basic idea is that when two senders transmit simultaneously, their packets superpose at the receiver, so that if the receiver already knows the content of one of the packets, it can cancel the interference and decode the other packet.
\DGnew{Not only can RODD take advantage of such known-interference cancellation techniques, it is also suitable for many-to-many communication, whereas it is difficult to perform interference cancellation for more than two users.}


\subsection{Decoding Delay}

The decoding delay of RODD is fixed to one frame interval, which is typically a few hundred symbols.  As shown through simulations, the delay of random access schemes for achieving the same error rate is many times larger than that of RODD.  Admittedly, the delay of random access can be as short as a short frame (tens of symbols), if luckily no other nodes happen to transmit at the same time; but the short frame is very likely to be lost due to collision, and many retransmissions are needed to achieve a desired performance.  RODD has an advantage if a fixed small delay is more desirable than a variable delay that has much larger expected value.

\subsection{Computational Complexity}

The computational complexity of the message-passing algorithm is linear in the frame length, the number of neighbors, and the number of messages each node can choose from.  RODD based on compressed sensing
with random signatures is most suitable for the situation where the broadcasts consist of a small number of bits. If each node has many bits to send, a structured code with low decoding complexity is needed for the scheme to be practical. One such code is the Reed-Muller code considered in~\cite{zhang2013neighbor}.

\subsection{Other Applications}
RODD can serve as a highly desirable sub-layer of any network protocol stack to provide the important function of simultaneous message exchange among neighbors. This sub-layer provides the missing link in many advanced resource allocation schemes, where it is often {\em assumed} that nodes are provided the state and/or demand of their neighbors.

Finally, the idea of using on-off signaling to achieve full-duplex communication using half-duplex radios applies to general peer-to-peer networks, and is not limited to mutual broadcasting traffic focused on in this paper.

\appendices
\section{Derivation of Message Computation~\eqref{eq:iter}} 
\label{a:bp}

We derive~\eqref{eq:iter} from~\eqref{eq:vv}. Denote $\Delta_{\mu k}=\sum_{j \in \partial \mu\backslash k}s_{\mu j}x_j$. The key to the simplification is to recognize that $\Delta_{\mu k}$ is approximately Gaussian. To be precise, if $\{x_j\}_{j \in \partial \mu\backslash k}$ were independent (conditioned on the observations traversed so far on the graph), then, by central limit theorem, $\Delta_{\mu k}$ converges weakly to a Gaussian random variable, whose mean is
\begin{equation}
v_{\mu k}^t = \sum_{j \in \partial \mu \backslash k}s_{\mu j}m_{j\mu}^t
\end{equation}
and variance is
\begin{equation}
(\sigma_{\mu k}^t)^2 = \sum_{j \in \partial \mu \backslash k}s_{\mu j}^2(\sigma_{j\mu}^t)^2 \,.
\end{equation}
Using the preceding Gaussian approximation,~\eqref{eq:U} can be calculated by a change of probability measure as
\begin{align}
U_{\mu k}^t(x) \propto &\int_{-\infty}^\infty
  \exp\left[-\left(y_{\mu}-\sqrt{\gamma_s}s_{\mu k}x -\sqrt{\gamma_s}\Delta \right)^2 \right] \nonumber \\
  &\cdot \frac{1}{\sqrt{2\pi(\sigma_{\mu k}^t)^2}} \exp\left[-\frac{1}{2(\sigma_{\mu k}^t)^2} \left(\Delta-v_{\mu k}\right)^2 \right] \diff \Delta \nonumber \\
  \propto &\exp\left[-\frac{1}{2\tau_{\mu k}^t}(x-z_{\mu k}^t)^2\right]
\end{align}
where $z_{\mu k}^t$ is defined in~\eqref{eq:z} and
\begin{equation}
\tau_{\mu k}^t = \sum_{j \in \partial \mu \backslash k}(\sigma_{j\mu}^t)^2+\frac{1}{2\gamma_ss_{\mu k}^2} \,.
\end{equation}
Using law of large numbers, we further approximate $\tau_{\mu k}^t$ by its average over all $(\mu,k)$ pairs with $s_{\mu k}\ne 0$, i.e., $\tau_{\mu k}^t$ is replaced by
\begin{align}
\tau^t &= \frac{1}{\sum_{\mu}|\partial \mu|}\sum_k\sum_{\mu\in k} \sum_{j\in\partial\mu\backslash k}(\sigma_{j\mu}^t)^2+\frac{1}{2\gamma_ss_{\mu k}^2} \nonumber \\
&\approx \frac{1}{\sum_\mu|\partial \mu|}\sum_\mu |\partial \mu|\sum_{j \in \partial \mu}(\sigma_{j\mu}^t)^2+\frac{1}{2\gamma_s}M_sq(1-q)
\end{align}
as shown in~\eqref{eq:tau}.

Now we have $U_{\mu k}^t \sim \mathcal{N}(z_{\mu k}^t,\tau^t)$, so it is easy to see that
\begin{equation}
\prod_{\nu \in \partial k \backslash \mu}U_{\nu k}^t \sim \mathcal{N}\left(\frac{\sum_{\nu \in \partial k \backslash \mu}z_{\nu k}^t}{|\partial k|-1},\frac{\tau^t}{|\partial k|-1}\right) \,.
\end{equation}
According to~\eqref{eq:V}, $V_{k\mu}^{t+1}(x)$ can be viewed as the conditional distribution $p_{X|Y}\left(x\,\Big|\,\frac{\sum_{\nu \in \partial k \backslash \mu}z_{\nu k}^t}{|\partial k|-1}\right)$, where $Y$ is the output of a Gaussian channel with noise $W\sim\mathcal{N}\left(0,\frac{\tau^t}{|\partial k|-1}\right)$. Therefore, by definition, $m_{k\mu}^{t+1}$ and $(\sigma_{k\mu}^{t+1})^2$ can be expressed as the conditional mean and conditional variance as in~\eqref{eq:condMean} and~\eqref{eq:condVar}.

\section{Proof of Proposition~\ref{prop:Pelb}} \label{a:aloha}
Let $\hat{\Phi}_1^p$ be an independent thinning of $\hat{\Phi}_1$ with retention probability $p$ to represent the transmitting nodes. It is easy to see that $\hat{\Phi}_1^p$ is an independent marked p.p.p. with intensity $\lambda p$. Denote the interference by
\begin{align}
I=\sum_{(Z_i,\mathcal{G}_i) \in \hat{\Phi}_1^p}G_{0i}|Z_i|^{-\alpha},
\end{align}
then we have
\begin{align}
\expect{\Probt{a}{s}(Z,\mathcal{G},\hat{\Phi}_1)} &= \expect{\expect{\ind{\frac{\gamma G|Z|^{-\alpha}}{\gamma I+1} \geq \delta}\,\Bigg|\,\hat{\Phi}_1^p}} \nonumber \\
&= \expect{\Prob\left\{G|Z|^{-\alpha}\geq \delta\left(I+\frac{1}{\gamma}\right)\Bigg|\hat{\Phi}_1^p\right\}}
\nonumber \\
&\LZnew{\leq} \expect{\left(\frac{\theta}{\delta}\right)^{\frac{2}{\alpha}} \left(I+\frac{1}{\gamma}\right)^{-\frac{2}{\alpha}}} \label{eq:exp_ps_ub} \\
&= \left(\frac{\theta}{\delta}\right)^{\frac{2}{\alpha}} \int_{-\infty}^\infty \left| i+\frac{1}{\gamma}\right|^{-\frac{2}{\alpha}}p_I(i)\diff i \label{eq:exp_ps_int}
\end{align}
where~\eqref{eq:exp_ps_ub} is derived from~\eqref{eq:ucdf} and $p_I$ is the pdf of 
 $I$.

\DGnew{Using the Laplace transform of $p_I$ given in~\cite{BacBla09FT},}
the Fourier transform\footnote{The reasons to work with Fourier transform in lieu of Laplace transform are: 1) The inverse Fourier transform here is easier to calculate; 2) the Fourier transform of $|i+\frac{1}{\gamma}|^{-\frac{2}{\alpha}}$ has a closed form.} of $p_I$
is obtained as
\begin{equation} \label{eq:Fourier1}
\mathcal{F}_{p_I}(\omega) = \exp\left\{-\lambda p(\iota\omega)^{2/\alpha}\frac{2\pi^2}{\alpha\sin(2\pi/\alpha)}\right\}\,.
\end{equation}
Since the Fourier transform of $|x|^a$ for $-1<a<0$ is
\begin{equation}
\mathcal{F}_{|x|^a}(\omega) = -\frac{2\sin(a\pi/2)\Gamma(a+1)}{|\omega|^{a+1}}\,,
\end{equation}
the Fourier transform of
\begin{align}
q_I(i)=\left|i+\frac{1}{\gamma}\right|^{-\frac{2}{\alpha}}
\end{align}
for $\alpha>2$ can be expressed as
\begin{equation} \label{eq:Fourier2}
\mathcal{F}_{q_I}(\omega) = e^{\iota\omega/\gamma}\frac{2\sin(\pi/\alpha)\Gamma(1-2/\alpha)}{|\omega|^{1-2/\alpha}} \,.
\end{equation}
Since the integral in~\eqref{eq:exp_ps_int} can be viewed as the Fourier transform of $p_I(i)q_I(i)$ at $\omega=0$, it can be calculated as the convolution of $\mathcal{F}_{p_I}(\omega)$ and $\mathcal{F}_{q_I}(\omega)$ at $\omega=0$~\cite{Oppenheim96}. Therefore, by~\eqref{eq:Fourier1} and~\eqref{eq:Fourier2}, we have
\begin{equation} \label{eq:int}
\int_{-\infty}^\infty \left|i+\frac{1}{\gamma}\right|^{-\frac{2}{\alpha}}p_I(i)\diff i = \frac{1}{2\pi} \mathcal{F}_{p_I}(\omega) \ast \mathcal{F}_{q_I}(\omega) \Bigg|_{\omega=0}
\end{equation}
where $\ast$ is the convolution operator. Therefore, according to~\eqref{eq:pe_lb},~\eqref{eq:exp_ps_int} and~\eqref{eq:int}, the error probability $\Probt{a}{e}$ can be lower bounded as
\begin{align}
\Probt{a}{e} \geq \Bigg(\max\Bigg\{0,1-&\frac{1}{2\pi} p(1-p) \left(\frac{\theta}{\LZnew{\delta}}\right)^{\frac{2}{\alpha}} \nonumber \\ &\mathcal{F}_{p_I}(\omega) \ast \mathcal{F}_{q_I}(\omega)\bigg|_{\omega=0} \Bigg\}\Bigg)^{\LZ{N_r}}. \label{eq:Pe_a}
\end{align}
Therefore,~\eqref{eq:aloha} in Proposition~\ref{prop:Pelb} follows
from~\eqref{eq:Pe_a} and~\eqref{eq:UBframes}.

\section{Proof of Proposition~\ref{prop:Pelb2}} \label{a:csma}
\LZH{For any $(Z_i,\mathcal{G}_i,T_i)\in\hat{\Phi}_1$, denote $G_i$ as the fading coefficient from node $Z_i$ to node $Z$.}
Define the following indicators for node $X$
\begin{align}
F_1 &= \ind{T_0>T} \\
F_2 &= \ind{T_i>T,\ \forall \LZH{\ (Z_i,\mathcal{G}_i,T_i)\in\hat{\Phi}_1} 
\ \text{with}\  G_i|Z_i-Z|^{-\alpha}\geq\theta} \\
F_3 &= \ind{\gamma G|Z|^{-\alpha}\geq\delta}
\end{align}
where $F_1=1$ if and only if the timer of $Z$ expires before that of $Z_0$, $F_2=1$ if and only if $Z$'s timer expires sooner than those of all its neighbors excluding $Z_0$, $F_3=1$ if and only if the received SNR from node $Z$ to node $Z_0$ exceeds the threshold $\delta$. In order for the transmission to be successful, we must have $F_1=F_2=F_3=1$. That is
\begin{align} \label{eq:Psub}
\expect{\Probt{c}{s}(Z,\mathcal{G},\hat{\Phi}_1)}
= \expect{F_1F_2F_3}.
\end{align}

Conditioned on $T=\varsigma$, we express the indicator \LZnew{$F_2$} as the value of some extremal shot-noise~\cite[Section~$2.4$]{BacBla09FT}. For fixed $\varsigma$, define the indicator of the event that $Z_i$ is a neighbor of $Z$ and it has a timer smaller than $\varsigma$:
\begin{align}
L(Z,Z_i,G_i,T_i) = \ind{G_i|Z_i-Z|^{-\alpha}\geq\theta\ \text{and}\ T_i<\varsigma}
\end{align}
for all $(Z_i,\mathcal{G}_i,T_i)\in\hat{\Phi}_1$. Define the extremal shot-noise at node $Z$ as
\begin{align}
\mathcal{Z}_{\hat{\Phi}_1}(Z) = \max_{(Z_i,\mathcal{G}_i,T_i)\in\hat{\Phi}_1} L(Z,Z_i,G_i,T_i).
\end{align}
Note that $\mathcal{Z}_{\hat{\Phi}_1}(Z)$ takes only two values $0$ or $1$ and consequently
\begin{equation}
\expect{F_2\Big|T=\varsigma} = \Prob\left\{\mathcal{Z}_{\hat{\Phi}_1}(Z)\leq 0\Big|T=\varsigma\right\}. \label{eq:expconde2}
\end{equation}
By~\cite[Proposition~$2.4.2$]{BacBla09FT},~\eqref{eq:expconde2} can be further calculated as
\begin{align}
&\expect{F_2\Big|T=\varsigma} \nonumber \\
&= \exp\left\{-\lambda\int_{\reals^2}\int_0^\infty\int_0^1 \ind{L(Z,z,g,t)=1}e^{-g} \diff t\diff g\diff z\right\} \nonumber \\
&= \exp\left\{-2\pi\lambda\varsigma\int_0^\infty\int_0^\infty \ind{gr^{-\alpha}\geq\theta}re^{-g} \diff g\diff r\right\} \nonumber \\
&= e^{-c\varsigma} \label{eq:e2expcond}
\end{align}
where $c$ is the average number of neighbors defined in~\eqref{eq:c}.

Therefore, according to~\eqref{eq:Psub}, we have
\begin{align}
\expect{\Probt{c}{s}(Z,\mathcal{G},\hat{\Phi}_1)}
&\leq \left(\frac{\theta\gamma}{\delta}\right)^{\frac{2}{\alpha}} \int_0^1(1-\varsigma)e^{-c\varsigma}\diff \varsigma \label{eq:exppsc} \\
&= \frac{1}{c^2} \left(\frac{\theta\gamma}{\delta}\right)^{\frac{2}{\alpha}} \left(e^{-c}+c-1\right) \label{eq:exppsc2}
\end{align}
where~\eqref{eq:exppsc} is derived from the the uniform distribution of $T_0$,~\eqref{eq:ucdf} and~\eqref{eq:expconde2}. Proposition~\ref{prop:Pelb2} then follows by combining~\eqref{eq:exppsc2} and~\eqref{eq:UBframes}.

As a by-product, by averaging over $\varsigma$ in~\eqref{eq:e2expcond}, which is uniformly distributed on $[0,1]$, the probability that a given node captures the channel to transmit in each slot can be calculated as $(1-e^{-c})/c$.

\bibliographystyle{ieeetr}
\bibliography{def,dguo,alljab,MutualBroadcast}

\end{document}